\documentclass[twocolumn,showpacs,preprintnumbers,amsmath,amssymb]{revtex4} 
\usepackage{hyperref}
\usepackage{amsfonts}
\usepackage{epsf}
\usepackage{t1enc}
\usepackage[latin2]{inputenc}

\usepackage{graphicx}
\usepackage{dcolumn}
\usepackage{bm}

\DeclareFontFamily{OT1}{rsfs}{} \DeclareFontShape{OT1}{rsfs}{m}{n}{
<-7> rsfs5 <7-10> rsfs7 <10-> rsfs10}{}
\DeclareMathAlphabet{\mycal}{OT1}{rsfs}{m}{n}

\def\scri{{\mycal I}}
\def\scrip{\scri^{+}}%

\begin{document}

\title{ What does a strongly excited 't\,Hooft-Polyakov magnetic
  monopole do? } 

\author{Gyula Fodor}
\author{Istv\'{a}n R\'{a}cz}

\affiliation{%
MTA KFKI, Részecske- és Magfizikai Kutatóintézet\\
H-1121 Budapest, Konkoly Thege Miklós út 29-33.\\
Hungary\\}%

\date{\today}

\begin{abstract}{\footnotesize
The time evolution of strongly exited $SU(2)$ Bogomolny-Prasad-Sommerfield
(BPS) magnetic monopoles in  Minkowski spacetime is investigated by
means of numerical simulations based on the technique of conformal
compactification and on the use of hyperboloidal initial value
problem. It is found that an initially static monopole does not
radiate the entire energy of the exciting pulse toward future null
infinity. Rather, a long-lasting quasi-stable `breathing state'
develops in the central region and certain expanding shell structures
-- built up by very high frequency oscillations -- are formed in the
far away region. }
\end{abstract}

\maketitle

\parskip 5pt

\small

The investigation of solitons in particle physics is of fundamental
interest (see e.g. \cite{rub} for a recent review). In particular,
considerable attention  has been paid to the  study of
't\,Hooft-Polyakov magnetic monopole solutions of coupled 
Yang-Mills--Higgs  (YMH) systems \cite{tH,P}. However, the relevant
investigations have been almost exclusively restricted to time
independent configurations. Thereby it is of obvious physical interest
to study dynamical properties of magnetic monopole configurations.

This paper is to report about the results of investigations concerning
the time evolution of a strongly excited spherically symmetric $SU(2)$  BPS
magnetic monopole \cite{ps,bo} on a fixed Minkowski background
spacetime by means of numerical techniques. In particular, the
underlying Yang-Mills--Higgs system is chosen so that the Yang-Mills field is
massive while the Higgs field is massless.
The dynamics starts by
hitting the static monopole by a concentrated pulse. First the
original pulse splits into two, a direct outgoing one and another one
going through the origin. Both pulses travel along null geodesics
taking away part of the energy of the excitation toward future null
infinity, $\scrip$, with the help of the massless Higgs field. It is
found, however, that this way the excited monopole releases only about
half  of the energy received.  The rest of the
energy of the original pulse seems to be restrained by the monopole in
accordance of which it develops a long lasting quasi-stable `breathing
state' in the central region and certain expanding shell structures in
the far away region. The frequency characterizing the breathing state
varies in  time and
it approaches asymptotically the value of the vector boson mass from
below. In  the far away region, where the Yang-Mills and the Higgs
fields are practically decoupled, the massless Higgs field does the
boring job of transporting the energy released gradually by the
central  monopole to $\scrip$, while the behavior of the massive
Yang-Mills field in the far away asymptotic region can
be characterized by the formation of certain expanding shell structures
where all the shells are built up by very high frequency
oscillations. These oscillations are found to be modulated by the product
of a simple time decaying factor of the form $t^{-{1}/{2}}$ and of an
essentially self-similar expansion.

The time decay of the examined quantities shows certain type of
{\it universality}. 
The total energy associated with the hyperboloidal hypersurfaces decays
in time with power $-2/3$, while  
the amplitude of the oscillating fields decay with power $-5/6$. 
In a recent work by Forg\'acs and 
Volkov \cite{fv}, based on the use of a linear
approximation of the BPS monopole, explanation is provided for
these universalities. 


The investigated dynamical magnetic monopole is described as a
coupled $SU(2)$  YMH system. The Yang-Mills field is represented by an
$\mathfrak{su}(2)$-valued vector potential $A_a$ and the associated 2-form
field $F_{ab}$ reads as
\begin{equation}
F_{ab}=\nabla_aA_b-\nabla_bA_a+i g\left[A_a,A_b\right]\label{ymf}
\end{equation}
where $[\ ,\ ]$ denotes the product in $\mathfrak{su}(2)$ and $g$
stands for the gauge coupling constant. The Higgs field (in the
adjoint representation) is given by an $\mathfrak{su}(2)$-valued
function $\psi$ while its gauge covariant derivative reads as
$\mathcal{D}_a\psi= \nabla_a\psi+i g[A_a,\psi]$. The dynamics of the
investigated YMH system is determined by the action
\begin{equation}
S=\int \left[Tr(F_{ef}F^{ef})
+2Tr(\mathcal{D}_e\psi\mathcal{D}^e\psi)\right]{\epsilon},
\end{equation}
where $\epsilon$ is the 4-dimensional volume element.

Our considerations were restricted to  spherically symmetric
configurations yielded by the `minimal' dynamical generalization of
the static 't\,Hooft-Polyakov magnetic monopole configurations
\cite{tH,P} (see also \cite{Hua}). Accordingly, the evolution took
place on Minkowski spacetime the line element of which, in spherical
coordinates $(t,r,\theta,\phi)$, is
$ds^{2}=dt^{2}-dr^{2}-r^{2}\left(  d\theta^{2}+\sin^{2}\theta\,d\phi
^{2}\right)$,  
while the Yang-Mills and Higgs fields, in the so called {\it abelian
gauge}, were assumed to posses the form   
\begin{eqnarray}
\hskip -.25cm A_a=-\frac{1}{g}\left[w\left\{\tau_{_{2}}(d\theta)_a-
\tau_{_{1}}\sin\theta(d\phi)_a \right\} +
\tau_{_{3}}\cos\theta(d\phi)_a\right] \\
\psi=H \tau_{_{3}},\phantom{+\left[\tau_{_{3}}\cos\theta(d\phi)_a\right]} 
\end{eqnarray}
where the generators $\{\tau_{_{I}}\}$ (I=1,2,3) of $\mathfrak{su}(2)$
are related to the Pauli matrices $\sigma_{_{I}}$ as $\tau_{_{I}}=
\frac12\sigma_{_{I}}$, moreover, $w$ and $H$ were assumed to be smooth
functions of $t$ and $r$.   

The field equations relevant for this system are 
\begin{eqnarray}
& & r^{2}{\partial ^{2}_r}{w}- r^{2}{\partial ^{2}_t}{w} =
w\left[\left({w}^{2}-1\right)+g^2{r^{2}}{H}^2 \right] 
\label{ymhe22}  \\ 
& &\hskip 0.35cm r^{2}{\partial ^{2}_r}{H}+ 2r{\partial_r }{H}
- r^{2}{\partial ^{2}_t}H =
2{w}^{2}H. \label{ymhe11} 
\end{eqnarray}
The only known analytic solution to (\ref{ymhe22}) and (\ref{ymhe11})
is the static BPS monopole  \cite{ps,bo}    
\begin{equation}
w_0 = \frac{gC r}{{\rm sinh}(gCr)},\ \ \ \ \ \ \ \
H_0 = C\left[\frac{1}{{\rm tanh}(gCr)}-\frac{1}{gC r}\right],\label{BPS}
\end{equation} 
where $C$ is an arbitrary positive constant. Since this solution is
stable it was
used first to check the efficiency of our numerical code. Later we 
considered the complete non-linear evolution of a system yielded by
strong impulse type excitations of this monopole. All the results
below concerns the evolution of such an excited BPS monopole.

The only scale parameter of the above described system is the vector
boson mass $m_w=gH\hspace{-.08cm}{}_{_\infty}$, where the limit value
$H\hspace{-.08cm}{}_{_\infty}=\lim_{r\rightarrow \infty}H$ of the
Higgs field can in general be shown
to be time independent \cite{fr2}. Since in the case considered here $
H\hspace{-.08cm}{}_{_\infty}=C\not=0$,
without loss of generality, the parameter choice
$g=H\hspace{-.08cm}{}_{_\infty}=m_w=1$ can be ensured to be satisfied
by making use of standard 
rescalings.  
 


To have a computational grid covering the full physical spacetime --
ensuring thereby that the outer grid boundary will not have an
effect on the time evolution -- the technique of conformal
compactification, along with the hyperboloidal initial value problem,
was used. This way it was possible to study the asymptotic
behavior of the fields close to future null infinity, as well as, the
inner region for considerably long physical time intervals.

The conformal transformation we used is a slight modification of
the static hyperboloidal conformal transformation applied by Moncrief
\cite{mon}. It is defined by introducing first the new
coordinates $T$ and $R$ instead of $t$ and $r$ as
\begin{equation}
T=\omega t-\sqrt{\omega^2 r^2+1}
\ \ \ {\rm  and} \ \ \  
R=\frac{\sqrt{\omega^2 r^2+1}-1}{\omega r},
\end{equation}
where $\omega$ is an arbitrary positive constant. The Minkowski
spacetime is covered by the coordinate domain given by the
inequalities  $-\infty < T < +\infty $ and $0 \leq R < 1$.
Then the conformally rescaled metric can be given as $\widetilde{g}_{ab}
=\Omega^2{g}_{ab}$, 
where the conformal factor is $\Omega={\omega}(1-R^2)/{2}$.
The $R=1$ coordinate line represents
$\scri^+$ through which the conformally rescaled metric
$\widetilde{g}_{ab}$ smoothly extends
to the coordinate domain with $R>1$.

Using the substitution
$H(t,r)={h(t,r)}/{r}+H\hspace{-.08cm}{}_{_\infty}$ 
the field equations, (\ref{ymhe22}) and (\ref{ymhe11}), in the new
coordinates read as
\begin{eqnarray}
&& {\mathfrak P}{ w}
=  w\left[\left({ w}^{2}-1\right)+
g^2\left( h+H\hspace{-.08cm}{}_{_\infty}
R\Omega^{-1}\right)^2\right] \label{2ymhe24}\\ &&
\phantom{{\mathfrak P}{\mathfrak P}{\widetilde w}}  
{\mathfrak P}{ h}=
2\left({h}+H\hspace{-.08cm}{}_{_\infty}{R}{\Omega}^{-1} \right) w^{2},
\label{2ymhe14}   
\end{eqnarray}
where the differential operator $\mathfrak P$ is defined as 
\begin{eqnarray}
\hskip -.35cm {\mathfrak P} &=&
\frac{4R^2}{(R^2+1)^2}\left[\frac{\Omega^2}{\omega^2}{\partial 
^{2}_R} - {\partial ^{2}_T} - 
2R{\partial_R\partial_T}\right. \nonumber \\ & &
\phantom{\frac{4R^2}{(R^2+)}} 
\left.-\frac{2\Omega}{\omega(R^2+1)}{\partial_T} - 
\frac{\Omega R(R^2+3)}{\omega(R^2+1)} 
{\partial_R}\right]. 
\end{eqnarray}
These equations can be put into the form of a first order strongly
hyperbolic system \cite{fr2}. The initial value problem for such a
system is known to be well-posed \cite{gko}. In particular, we solved
this first order system 
numerically by making  use of the `method of line' in a fourth
order Runge-Kutta scheme following the recipes proposed by Gustafsson
{\it et al} \cite{gko}. All the details related to the numerical
approach, including representations of derivatives, treatment
of the grid boundaries are to be published in
\cite{fr2}. The convergence tests justified that our code 
provides a fourth order representation of the selected evolution
equations. Moreover, the monitoring of the energy conservation and the
preservation of the constraint equations, along with the coincidence
between the field values which can be deduced by making use of the
Green's function and by the adaptation of our numerical code to the
case of massive Klein-Gordon fields, made it apparent that the
phenomena described below have to be, in fact, physical properties of the
magnetic monopoles.

In each of the numerical simulations initial data on the $T=0$
hypersurface was specified for the system of our first order
evolution equations. In particular, a superposition of the 
data associated with the BPS monopole, (\ref{BPS}),
and of an additional pulse of the form 
\begin{equation} \hskip -.27cm(\partial_T w)_\circ = \left\{
\begin{array} {rr}  
    c \exp\left[\frac{d}{(r-a)^2-b^2}\right], & {\rm if}\ r\in
    [a-b,a+b] ;\\  0 , & {\rm otherwise},  \end{array} \right.\label{ff}
\end{equation}
with $a \geq b>0$, which is a smooth function of compact support, was
used. This 
choice, providing non-zero time derivative for $w$, corresponds
to ``hitting'' the static monopole configuration 
between two concentric shells at $r=a-b$ and $r=a+b$ with a bell shape
distribution.  Basically the same type of evolution occurs when
instead  of  $(\partial_T w)_\circ$ we prescribe $(\partial_T
h)_\circ$ in a similar fashion. 

For the sake of brevity, all the simulations shown below refer to the
same pulse (\ref{ff}) corresponding to the choice of the parameters 
$a=2$, $b=1.5$, $c=70$, $d=10$ and $\omega=0.05$. The energy of this
pulse is $55.4\%$ of the static 
monopole. Hence, the yielded dynamical system cannot be considered as
being merely a simple perturbation of the static monopole. We would
also like to emphasize that the figures shown below are typical
in the sense that for a wide range of the parameters characterizing
the exciting pulse, qualitatively, and in certain cases even
quantitatively, the same type of responses are produced by the
monopole \cite{fr2}.

\begin{figure}[htbp!]
 \centerline{
  \epsfxsize=8.5cm 
\epsfbox{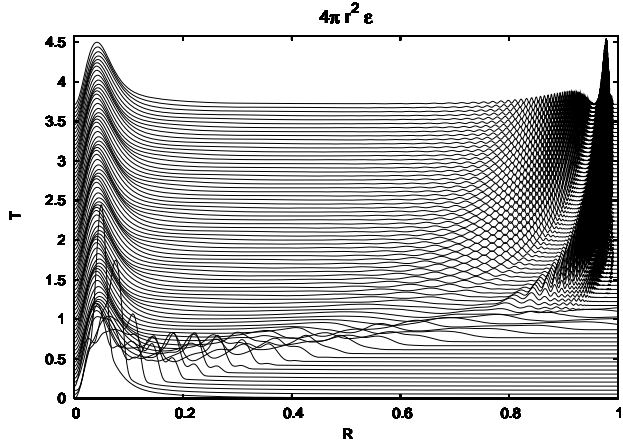}
 }
\caption{\footnotesize \label{figT00s}Spacetime diagram showing the
 time evolution of the energy density associated with shells of radius
 $r$, i.e. of $4\pi r^2\varepsilon$.}
\end{figure}

To start off consider first the product of the energy density
$\varepsilon$ and $4\pi r^2$ plotted 
on succeeding $T=const$ hypersurfaces, providing thereby a
spacetime picture of the time evolution (see Fig.\,\ref{figT00s}). 
The use of $4\pi r^2\varepsilon$, instead of $\varepsilon$,  makes it
easier to see the  main characteristics of the dynamics up to
$\scri^+$. In the early part the direct energy  transport to
$\scri^+$ by the Higgs field, with the velocity of light, is apparent.
The developing `breathing state' of the monopole and the formation of
the expanding shells of high  frequency oscillations are both clearly
manifested.

\begin{figure}[htbp!]
 \centerline{
  \epsfxsize=8.5cm 
\epsfbox{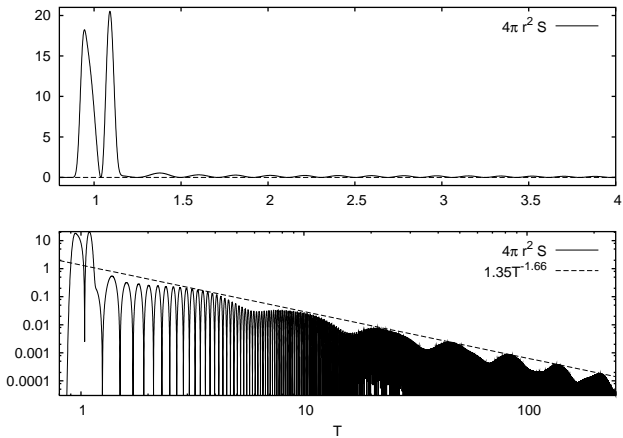}
 }
\caption{\footnotesize \label{figS} The time dependence of the
  product of the energy
current density $S$ and $4\pi r^2$ at $\scri^+$ is shown on the
intervals $0.8\leq T\leq 4$ and $0.8\leq T\leq 250$.}
\end{figure}

The energy radiated to $\scri^+$ can be pictured by plotting (see
Fig.\,\ref{figS}) the product of the energy current density $S$ and
$4\pi r^2$ against time $T$ at $\scri^+$ ($R=1$).
It is apparent that the arrival of the two pulses is followed by a
small scale but systematic energy loss of the system which has exactly
the same characteristic period as the inner breathing state of the
monopole. From the logarithmic plot the asymptotic behavior
$4\pi r^2 S_{asympt}\approx C_{_{S}} 
T^{-\gamma_{_{S}}}$ can be read off where $\gamma_{_{S}}\approx
1.66$. In virtue of the energy conservation this relation
implies that the energy associated with the $T=const$
hypersurfaces approaches to the energy of the asymptotic final state 
as $T^{-2/3}$,
in agreement with \cite{fv}.

\begin{figure}[htbp!]
 \centerline{
  \epsfxsize=8.5cm 
\epsfbox{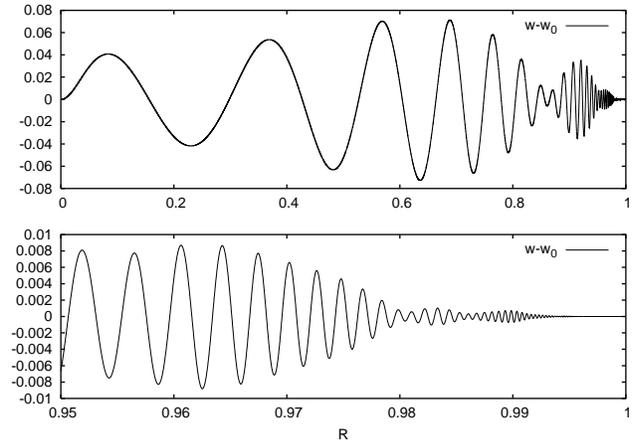}
 }
\caption{\footnotesize \label{figws} The difference $w-w_0$  is
  plotted on time slice $T=1.695$. The formation of the high frequency
  oscillations sorted in certain shell structures is transparent.} 
\end{figure}

By inspection of the evolution of the field variables $w$ and $h$ it
is obvious that the expanding high frequency oscillations are
associated with the massive Yang-Mills field
exclusively. Fig.\,\ref{figws} shows a constant time slice at
$T=1.695$ of the evolution of $w-w_0$.
The formation of the shells built up by high frequency oscillations is
transparent on Fig.\,\ref{figws}, which is reminiscent of Figs.\,1.\,and
2.\,of \cite{fr}. The behavior of these oscillations in the asymptotic
region can be explained by referring to results of \cite{fr} where
this phenomenon has already been found to characterize the evolution
of the simplest linear massive scalar field. In doing  this, notice
first that by (\ref{ymhe22}) $w/r$ satisfies asymptotically the
massive Klein-Gordon equation. This, along with the claims of
\cite{fr}, implies then that the oscillations are modulated by two
factors. By an overall factor $t^{-1/2}$ scaling down the oscillations
in time, moreover, by an essentially self-similar expansion, i.e. by a
function depending on $t$ and $r$ only in the combination $\rho=r/t$.

Probably, the most interesting unexpected feature of the time evolution is the
appearance of the breathing state of the monopole (see the central
region of  Fig.\,\ref{figT00s}). To have a quantitative
characterization of this phenomenon it is informative to consider a
constant $R$ slice of the deviation $\varepsilon-\varepsilon_0$ of the
full energy density $\varepsilon$ and that of the static monopole
$\varepsilon_0$.
\begin{figure}[htbp!]
 \centerline{
  \epsfxsize=8.5cm 
\epsfbox{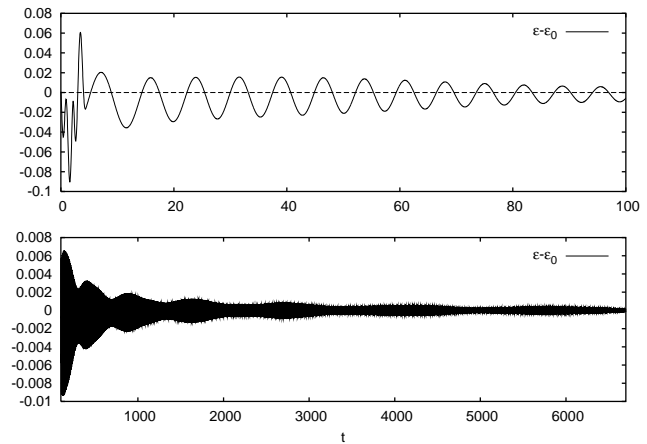}
 }
\caption{\footnotesize \label{figEpsR} The deviation
$\varepsilon-\varepsilon_0$ of the full energy density $\varepsilon$
from that of the static monopole $\varepsilon_0$ is plotted against the
physical time $t$ for intervals $0\leq t\leq 100$ and $100\leq t\leq
6700$ at $R=0.0254$. It is 
striking how quickly the breathing state develops.} 
\end{figure}   
The breathing state of the monopole can be characterized by the time
dependence of the amplitude and the frequency of the associated
quasi-normal oscillations, as well as, by the power spectrum of the
oscillations. The  time dependence of the amplitude is shown by
Fig.\,\ref{figEpsR}. 
Note that the center of the oscillations is actually lower than
$\varepsilon_0$, which implies that the time average of the energy 
contained in the central region is smaller than the energy contained
in the same region of the static 
monopole. This is most likely due to nonlinear effects.

\begin{figure}[htbp!]
 \centerline{
  \epsfxsize=8.1cm 
\epsfbox{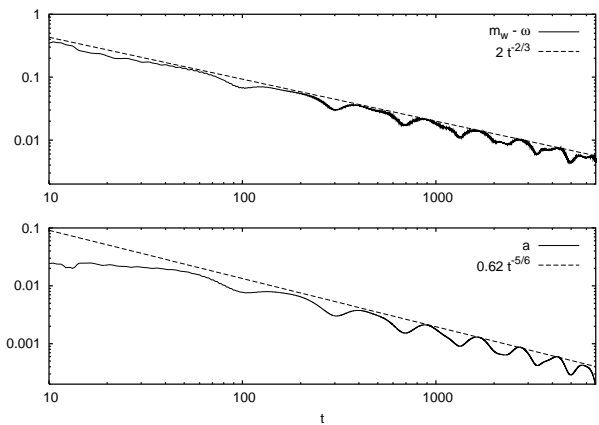}
 }
\caption{\footnotesize \label{figwt} The frequency difference $m_w-\omega$
and the amplitude $a$ of the
oscillations of the breathing state, shown by Fig.\,\ref{figEpsR}, is
plotted against the 
physical time $t$. }
\end{figure}

The time dependence of the frequency and the amplitude of the 
oscillations was determined by fitting a simple function of the form 
$\varepsilon-\varepsilon_0 = a \sin(\omega t +b) +c$ to the numerical 
data on a sufficiently short time interval so that this interval was 
shifted point by point through the entire time evolution. 
The resulted graph is shown by
Fig.\,\ref{figwt}.
The frequency of the oscillations
is essentially increasing, asymptotically approaching the value of the
vector boson mass $m_w$. 
The logarithmic plot suggests a relatively
simple form for the overall behavior of the frequency-time
function. In particular, 
the approximate relation $\omega \approx m_w - C_{_\omega}
t^{-\gamma_{_{\omega}}}$ seems to be valid not merely asymptotically
but for the entire evolution, where ${\gamma_{_{\omega}}}$ was found
up to a high accuracy to be $2/3$. Exactly the same value
${\gamma_{_{\omega}}}=2/3$ was found in \cite{fv}.
The asymptotic behavior of the amplitude of the
oscillations can be
approximated by the 
simple form  $a_{asympt} \approx 
C_a t^{-\gamma_{_{a}}}$, where
$\gamma_{_{a}} \approx 0.833$, which is in good agreement
with the value of $-5/6$ of \cite{fv}.
The energy contained in some finite radius is described by the time average
of the energy, i.e. by the function $c$, not by the amplitude $a$. 
For this reason there is no contradiction between the exponents $-5/6$ 
in the central region and $-2/3$ at infinity. The function $c$ appears
to be smaller and decays faster than the amplitude, and can also take 
positive and negative values depending on the location.
  
\begin{figure}[t]
 \centerline{
   \epsfxsize=8.1cm
\epsfbox{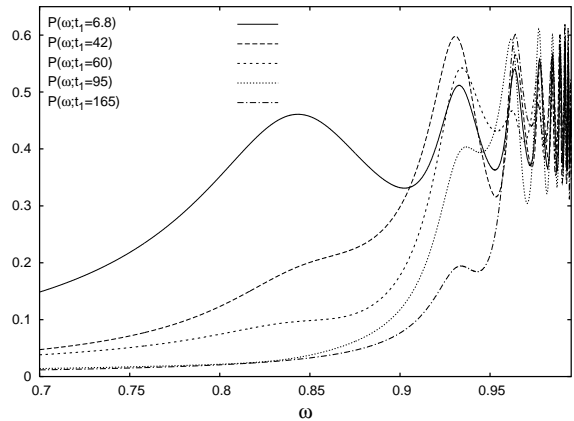}
 }
\caption{\footnotesize \label{figw} The power spectrum of the
oscillations of $\varepsilon-\varepsilon_0$ at $R=0.0254$ relevant for
the intervals $[t_1,t_2]$, with $t_2=11107$ while $t_1$
taking the values $6.8$, $42$, $60$, $95$ and $165$ is shown.} 
\end{figure}

It is also of interest to consider the power spectrum
$P(\omega;t_1,t_2)$ of 
the oscillations defined as twice of the absolute value of the Fourier
transform $\widehat{(\varepsilon-\varepsilon_0)}(\omega;t_1,t_2)=(2\pi)^{-1/2}
\int^{t_2}_{t_1} \left(\varepsilon-\varepsilon_0\right)(t) e^{-i \omega
t}dt$. Fig.\,\ref{figw} shows $P(\omega;t_1,t_2)$ with $t_2$ having
the fixed value $t_2=11107$ and $t_1$ being 
chosen to take the values $6.8$, $42$, $60$, $95$ and $165$,
respectively.  By varying $t_1$ it is possible to monitor the change
of the frequency 
of the oscillations in the relevant early period.  These graphs
support the perturbative result of Forg\'acs and 
Volkov \cite{fv} claiming that there has to be an infinite 
family of `resonant states', associated with the breathing state of the
monopole, possessing discrete frequencies. See, for instance, the
peaks of the graph 
of $P(\omega;t_1=6.8)$ at $\omega \approx 0.844,
0.933, 0.9645, $ $ 0.9775, 0.985,...$ . 
According to Fig.\,\ref{figw} the
lower frequency members of the resonant states die out faster than the
higher frequency ones which is consistent with the above mentioned
increase of the overall frequency of the breathing state. 

In summary, the time evolution of strongly excited BPS monopoles has
been studied. It is found to be generic that in the central region a
long lasting
quasi-periodic breathing state develops.
The behavior of all the examined 
quantities justifies the intuitive expectation that in the inner region 
the system settles down to the original static BPS
monopole, while the self-similarly expanding oscillations disperse
asymptotically in
the far away region.

\small

{\bf Acknowledgments} The authors wish to thank P\'{e}ter
Forg\'{a}cs who suggested the numerical investigation of magnetic
monopoles. We are indebted to him and to Michael Volkov 
for communicating to us their results prior to publication.
We also would like to thank them and J\"{o}rg
Frauendiener for discussions. 
This research was supported in parts by OTKA
grant T034337 and NATO grant PST.CLG. 978726. 

\vfill\eject
\end{document}